\documentclass[preprint,pra]{revtex4-1}

\usepackage{graphicx}
\usepackage{color}
\usepackage{amsmath,amssymb,bm,mathrsfs}

\newcommand{\lap}{\nabla^{2}}

\newcommand{\Br}{{\bf r}}

\newcommand{\Bx}{{\bf x}}
\newcommand{\By}{{\bf y}}

\newcommand{\Bk}{{\bf k}}

\renewcommand{\r}{{\bf \hat r}}

\newcommand{\khat}{{\bf \hat k}}
\newcommand{\x}{{\bf \hat x}}
\newcommand{\y}{{\bf \hat y}}

\newcommand{\ket}[1]{\left| #1 \right\rangle}
\newcommand{\bra}[1]{\left\langle #1 \right|}

\newcommand{\Epos}{\widehat E^{+}}
\newcommand{\Eneg}{\widehat E^{-}}
\newcommand{\E}{\widehat E}

\newcommand\h{{h_l^{(1)}}}

\begin{document}

\title{Scattering of entangled two-photon states}

\author{John C. Schotland}
\affiliation{Department of Mathematics and Department of Physics, University of Michigan, Ann Arbor, MI 48109}
\email{schotland@umich.edu}

\author{A. Caz\'e}
\affiliation{Department of Mathematics, University of Michigan, Ann Arbor, MI 48109}
\email{acaze@umich.edu}

\author{Theodore B. Norris}
\affiliation{Department of Electrical Engineering and Computer Science, University of Michigan, Ann Arbor, MI 48109}
\email{tnorris@umich.edu}

\date{\today}

\begin{abstract}
We consider the scattering of entangled two-photon states from collections of small particles. We also study the related Mie problem of scattering from a sphere. In both cases, we calculate the entropy of entanglement and investigate the influence of the entanglement of the incident field on the entanglement of the scattered field.
\end{abstract}
 
\maketitle

The propagation of quantum states of light in complex media such as the atmosphere, colloidal suspensions and biological tissue is a topic of fundamental interest and considerable applied importance. New physical phenomena, including the transmission of quantum states through random media~\cite{Lodahl_2005_1,Lodahl_2005_2,Cande_2014,Ott_2010}; the observation of spatial correlations in multiply-scattered squeezed light~\cite{Smolka_2009, Smolka_2012}; and the measurement of two-photon speckle patterns~\cite{Peeters_2010,Pires_2012,Beenakker_2009} have been described. Applications to spectroscopy~\cite{Skipetrov_2007}, imaging~\cite{Klyshko_1988,Strekalov_1995,Abouraddy_2001,Abouraddy_2004,Gatti_2004,Scarcelli_2004, Scarcelli_2006,Erkmen_2008,DAngelo_2005,Schotland_2010,Nasr_2003,Teich_2012} and communications~\cite{Moustakas_2000,Skipetrov_2008,Shapiro_2009,Yuan_2010,Tworzydlo_2002} have also been reported. To understand the underlying physical principles, it is often useful to consider relatively simple model systems. A step in this direction was taken in~\cite{Markel_2014}, where the propagation of two-photon entangled states in random media was studied within the framework of radiative transport theory.   
In this Letter, we investigate the scattering of entangled two-photon states in an even simpler setting, namely from \emph{deterministic} media. We study in detail the problem of scattering from collections of small particles, as well as the related Mie problem of scattering from a sphere.  In particular, we analyze the extent to which scattering can alter the quantum correlations of the optical field and calculate the resulting entropy of entanglement. 

We begin by  considering the propagation of a quantized field in a material medium with 
dielectric permittivity $\varepsilon$. For simplicity, we work with the scalar theory of the electromagnetic field. The electric-field operator $\E$ may be decomposed into positive and negative frequency components: $\E=\Epos + \Eneg$. The positive frequency component $\Epos$ obeys the wave equation~\cite{Glauber_1991,Scheel_1999}
\begin{equation}
\label{wave_eqn}
\lap \Epos = \frac{\varepsilon (\Br)}{c^2} \frac{\partial^2 \Epos}{\partial t^2} \ .
\end{equation}
The negative frequency component $\Eneg$ is defined by $\Eneg=[\Epos]^\dag$.
Here the medium is taken to be nonabsorbing, so that $\varepsilon$ is purely real and positive. Now, let $\ket{\psi}$ be a two-photon state and define the second-order coherence function $\Gamma^{(2)}$ as the normally ordered expectation of field operators:
\begin{eqnarray}
\nonumber
\Gamma^{(2)}(\Br,t;\Br',t') = \bra{\psi}\Eneg(\Br,t)\Eneg(\Br',t') \quad\quad \\
\times \Epos(\Br',t')\Epos(\Br,t)\ket{\psi}
\ . 
\end{eqnarray}

We note that $\Gamma^{(2)}$ is proportional to the probability of detecting one photon at $\Br$ at time $t$ and a second photon at $\Br'$ at time $t'$. It can be measured in a Hanbury-Brown--Twiss interferometer~\cite{Mandel_Wolf}. It can be seen that $\Gamma^{(2)}$ factorizes~\cite{Rubin_1996} as
\begin{eqnarray}
\Gamma^{(2)}(\Br,t;\Br',t') = |A(\Br,t;\Br',t')|^2 \ .
\end{eqnarray}
Here $\ket{0}$ is the vacuum state and the two-photon amplitude $A$ is defined by
\begin{equation}
A(\Br,t;\Br',t') = \bra{0}\Epos(\Br,t)\Epos(\Br',t')\ket{\psi} \ .
\end{equation}
Evidently, $A$ satisfies the pair of wave equations
\begin{eqnarray}
\label{phi_1}
\lap_{\Br}A &=& \frac{\varepsilon(\Br)}{c^2}\frac{\partial^2 A}{\partial t^2} \ , \\
\lap_{\Br'}A &=& \frac{\varepsilon(\Br')}{c^2}\frac{\partial^2 A}{\partial t'^2} \ ,
\end{eqnarray}
which  follow from the fact that $\Epos$ obeys the wave equation~\cite{Saleh_2005}. 
We will find it convenient to introduce the Fourier transform of the two-photon amplitude $A$, which is defined by
\begin{equation}
\widetilde A(\Br,\omega;\Br',\omega')= \int dt dt' e^{i(\omega t + \omega' t')} A(\Br,t;\Br',t') \ .
\end{equation}
Eq.~(\ref{phi_1}) then becomes
\begin{eqnarray}
\label{phi_1_helmholtz}
\lap_{\Br} \widetilde A + k^2\varepsilon(\Br) \widetilde A = 0 \ , \\
\lap_{\Br'} \widetilde A + k'^2\varepsilon(\Br') \widetilde A = 0 \ ,
\end{eqnarray}
where $k=\omega/c$ and $k'=\omega'/c$. 

\begin{figure}[t] 
\vspace{-0.5in}    
\centering
\includegraphics[width=4.in]{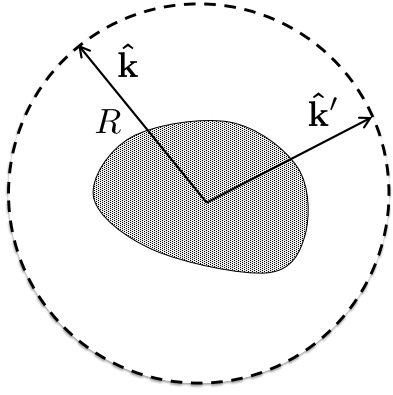}
\vspace{-0.5in}
\caption{Illustrating the notation.}
\label{fig1}
\end{figure}

We now develop the scattering theory for the two-photon amplitude.
To proceed, we consider the Helmholtz equation
\begin{equation}
\label{helmholtz}
\lap u + k^2\varepsilon(\Br) u = 0 \ .
\end{equation}
The field $u$ is taken to consist of incident and scattered parts, which we write as the sum $u=u_i+u_s$.
The incident field $u_i$ is the field that would exist in the absence of the scatterer.
The scattered field $u_s$ is given by~\cite{economou_book}
\begin{equation}
\label{u_s}
u_s(\Br)=\int d^3r_1 d^3r_2 G(\Br,\Br_1)T(\Br_1,\Br_2)u_i(\Br_2) \ ,
\end{equation}
where the Green's function $G$ is of the form
\begin{equation}
\label{Green}
G(\Br,\Br') = \frac{e^{ik|\Br-\Br'|}}{|\Br-\Br'|} \ .
\end{equation}
The $T$-matrix obeys the integral equation 
\begin{equation}
\label{T-matrix}
T(\Br,\Br') = k^2\eta(\Br)\delta(\Br-\Br') + k^2\eta(\Br)\int d^3r'' G(\Br,\Br'')T(\Br'',\Br') \ ,
\end{equation}
where the susceptiblity $\eta$ is defined by the relation $\varepsilon=1+4\pi\eta$.
Applying (\ref{u_s}) to each of the arguments of $\widetilde A$, we find that the Fourier transformed two-photon amplitude of the scattered field, denoted $A_s$, is given by
\begin{eqnarray}
\label{phi_s}
\nonumber
A_s(\Br,\Br')= \int d^3r_1 d^3r_2d^3r'_1 d^3r'_2G(\Br,\Br_1)T(\Br_1,\Br_2) \\
\times G'(\Br',\Br'_1)T'(\Br'_1,\Br'_2) A_i(\Br_2,\Br'_2) \ , \quad
\end{eqnarray}
where $A_i$ is the Fourier-transformed two-photon amplitude of the incident field. In addition, $G'$ and $T'$ denote the Green's function and $T$-matrix at the wavenumber $k'$. If $A_s(\Br,\Br')$ factorizes into a product of two functions which depend upon $\Br$ and $\Br'$ separately, we will say that the two-photon state $\ket{\psi}$ is not entangled. In contrast, an entangled state is not separable. We can now state our first result. It follows directly from (\ref{phi_s}) that if $A_i$
is separable then $A_s$ is separable. That is, if $A_s$ is entangled then $A_i$ is entangled, which means that entanglement cannot be created by scattering an unentangled incident state.

We now consider the far-field limit of the two-photon amplitude $A_s(\Br,\Br')$, where
the points of observation $\Br$ and $\Br'$ lie on a sphere of radius $R$ in the far-zone of the scatterer, as shown in Figure~1. In doing so, we make use of the asymptotic behavior of the Green's function $G(\Br,\Br')$ for $r \gg r'$:
\begin{equation}
G(\Br,\Br') \sim \frac{e^{ik r}}{r} e^{-ik\hat\Br\cdot\Br'} \ .
\end{equation}
We also use the plane-wave expansion for the two-photon amplitude $A_i$, which is of the form
\begin{equation}
A_i(\Br,\Br') = \int d\khat d\khat' {\mathcal A}_i(\hat\Bk,\hat\Bk') e^{i(k\hat\Bk\cdot\Br + k'\hat\Bk'\cdot\Br')}
 \ .
\end{equation}
Here ${\mathcal A}_i$ is a suitable coefficient, which is expressible as a vacuum to two-photon state transition amplitude. Thus (\ref{phi_s}) becomes
\begin{equation}
\label{psi_i-psi_s}
{\mathcal A}(\hat\Bk,\hat\Bk') =\int d \khat_1 d \khat_2 \bra{\Bk}T\ket{\Bk_1}\bra{\Bk'}T'\ket{\Bk_2}{\mathcal A}_i(\hat\Bk_1,\hat\Bk_2) \ ,
\end{equation}
where $\Bk = k\hat\Br$, $\Bk' = k'\hat\Br'$ and ${\mathcal A} = R^2\exp\left[-i(k+k')R\right]A_s$, with $|\Br|=|\Br'|=R$. See Figure~1 where the geometry of the problem is illustrated. We note that the above $T$-matrices are on-shell. The momentum-space $T$-matrix elements are defined by
\begin{eqnarray}
\bra{\Bk}T\ket{\Bk'}=\int d^3r d^3r' e^{-i(\Bk\cdot\Br-\Bk'\cdot\Br')}
T(\Br,\Br') \ ,
\end{eqnarray}
where $|\Bk|=|\Bk'|=k$. Given a scattering medium characterized by its $T$-matrix, (\ref{psi_i-psi_s}) predicts the two-photon amplitude of the far-zone scattered field in terms of the two-photon amplitude of the incident field. As may be expected, if ${\mathcal A}_i$ is separable then ${\mathcal A}$ is separable, consistent with (\ref{phi_s}). If ${\mathcal A}_i(\khat_1,\khat_2)=\delta(\khat_1-\khat_2)$, which corresponds to a fully entangled two-photon state, then (\ref{psi_i-psi_s}) becomes
\begin{equation}
\label{Psi_s-entangled}
{\mathcal A}(\hat\Bk,\hat\Bk') =\int d \khat''\bra{\Bk}T\ket{\Bk''}\bra{\Bk'}T'\ket{\Bk''} \ ,
\end{equation}
where $|\Bk|=k$ and $|\Bk'|=k'$.

We now compute ${\mathcal A}$ for several different scattering systems. We begin with a small spherical scatter of radius $a$, where $ka \ll 1$. 
The $T$-matrix is then given by
\begin{equation}
\label{point_scatterer}
\bra{\Bk}T\ket{\Bk'} = t(k)e^{i(\Bk-\Bk')\cdot\Br_0}  \ ,
\end{equation}
where $\Br_0$ is the position of the scatterer and $t(k)$ is defined in the supplementary material.
Making use of (\ref{Psi_s-entangled}) and (\ref{point_scatterer}) we obtain
\begin{eqnarray}
\nonumber
{\mathcal A}(\hat\Bk,\hat\Bk')=4\pi t(k)t(k')e^{i(k\khat + k' \khat')\cdot\Br_0}
{\rm sinc}\left(|(k+k')\Br_0|\right) \ . \\
\end{eqnarray}
We see at once that ${\mathcal A}$ is separable and thus the scattered field is unentangled, even when the incident field is entangled.

\begin{figure}[t]     
\centering
\includegraphics[width=3.in]{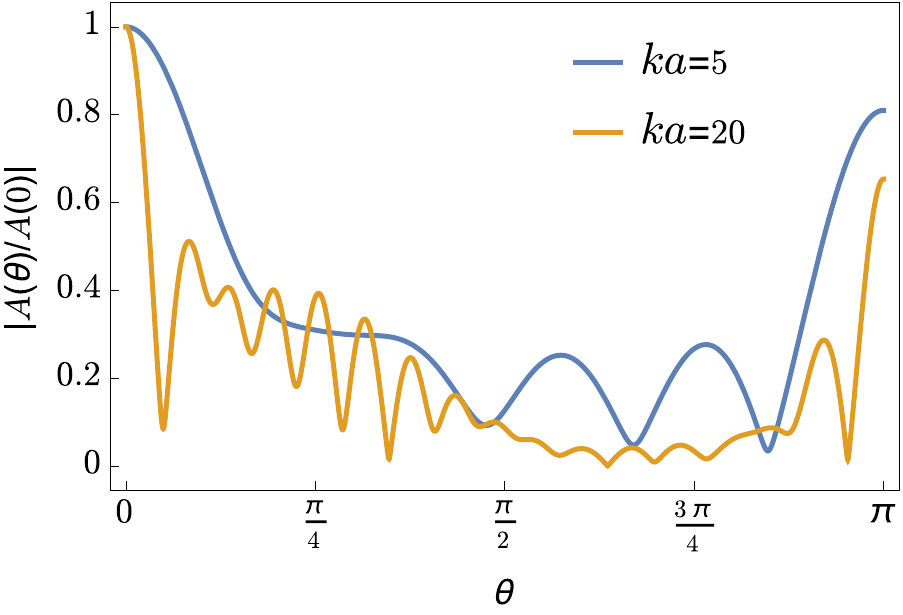} 
\caption{Two-photon amplitude ${\mathcal A}(\khat,\khat')$ of a spherical scatterer of radius $a$ as a function of the angle $\theta$ between $\khat$ and $\khat'$. The index of refraction of the sphere is $n=1.5$. }
\label{fig2}
\end{figure}

Next, we consider a collection of identical small scatterers. 
The $T$-matrix is given by
\begin{equation}
\label{T-multiple}
\bra{\Bk}T\ket{\Bk'} = \sum_{a,b}t_{ab}(k)e^{i(\Bk\cdot\Br_a-\Bk'\cdot\Br_b)} \ ,
\end{equation}
where $\{\Br_a\}$ are the positions of the scatterers and $t_{ab}$ is defined in the supplementary material. Using (\ref{Psi_s-entangled}), we find that ${\mathcal A}$ is given by
\begin{eqnarray}
\label{psi_s-particles}
\nonumber
{\mathcal A}(\hat\Bk,\hat\Bk') = 4\pi \sum_{a,b}\sum_{a',b'}t_{ab}(k)t_{a'b'}(k')
e^{i(k\khat\cdot\Br_a + k'\khat'\cdot\Br_{a'})} \\
\times {\rm sinc}(|k\Br_b + k'\Br_{b'}|) \ . 
\quad\quad\quad\quad\quad
\end{eqnarray}
We note that in general $A$ is nonseparable; thus the scattered field is entangled.

Finally, we consider a homogeneous spherical scatterer of radius $a$ centered at the origin with index of refraction $n$. The $T$-matrix is of the form
\begin{equation}
\bra{\Bk}T\ket{\Bk'} = \sum_l (2l+1)A_l(k) P_l(\khat\cdot\khat') \ ,
\end{equation}
where the Mie coefficient $A_l$ is defined as~\cite{Grandy}
\begin{equation}
A_l(k) = \frac{1}{ik}\frac{j_l(nka)j_l'(ka) - nj_l(ka)j_l'(nka)}
{n\h(ka)j_l'(nka)-\h'(ka)j_l(nka)} \ .
\end{equation}
Using the identity
\begin{equation}
\label{identity}
\int d\khat'' P_l(\khat\cdot\khat'')P_{l'}(\khat'\cdot\khat'') = \frac{4\pi}{2l+1}\delta_{ll'}P_l(\khat\cdot\khat')
\end{equation}
to carry out the integral in (\ref{Psi_s-entangled}), we find that ${\mathcal A}$ is given by
\begin{equation}
\label{psi_sphere}
{\mathcal A}(\khat,\khat') = 4\pi \sum_l (2l+1) A_l(k) A_l(k')P_l(\khat\cdot\khat') \ .
\end{equation}
In Figure~2 the quantity ${\mathcal A}(\khat,\khat')$ is plotted as a function of the angle between $\khat$ and $\khat'$.

We now turn to the computation of the entanglement entropy for the above systems. Following~\cite{Law_2004}, we consider the singular value decomposition (also known as the Schmidt decomposition) of the two-photon amplitude, viewed as an operator with kernel ${\mathcal A}(\khat,\khat')$. We find that ${\mathcal A}(\khat,\khat')$ can be decomposed into 
a superposition of separable terms of the form
\begin{equation}
{\mathcal A}(\khat,\khat') = \sum_n \sigma_n u_n(\khat) v_n^*(\khat') \ ,
\end{equation}
where each term can be interpreted as not entangled. 
Here the singular values $\sigma_n$ are real-valued and the singular functions
obey
\begin{eqnarray}
\label{def_un}
\int \left({\mathcal A}^*{\mathcal A}\right)(\khat,\khat')v_n(\khat')d\khat' &=& \sigma_n^2 v_n(\khat) \ , \\
\int \left({\mathcal A}{\mathcal A}^*\right)(\khat,\khat')u_n(\khat')d\khat' &=& \sigma_n^2 u_n(\khat) \ ,
\label{def_vn}
\end{eqnarray}
where ${\mathcal A}^*$ is the adjoint of the operator $\mathcal A$.
A measure of the degree of entanglement is the entropy $S$, which is defined by
\begin{equation}
\label{entropy}
S = -\sum_n \sigma_n \log \sigma_n \ .
\end{equation}
We note that the larger the value of $S$, the greater the degree of entanglement.

To illustrate the above results, we calculate the entanglement entropy of a spherical scatterer. Using (\ref{psi_sphere}) and the identity (\ref{identity}), we find that
\begin{equation}
\left({\mathcal A}^*{\mathcal A}\right)(\khat,\khat') = (4\pi)^4\sum_{l,m}\left(A_l(k)A_l(k')\right)^2 Y_{lm}(\khat)
Y_{lm}^*(\khat') \ .
\end{equation} 
We immediately see that the singular functions and singular values are given by
\begin{eqnarray}
u_{lm}(\khat) = v_{lm}(\khat) = Y_{lm}(\khat) \ , \\
\sigma_{lm} =  \sigma_l = (4\pi)^2 |A_l(k)A_l(k')| \ .
\end{eqnarray}
Thus the entropy is given by the formula
\begin{equation}
S = -\sum_l (2l+1)\sigma_l \log \sigma_l \ .
\end{equation} 
In Fig.~3 we plot the entropy as a function of the radius of the sphere. We see that in the limit where the radius tends to zero, the entropy vanishes, consistent with the separability of the two-photon amplitude for the case of a point scatterer in (\ref{point_scatterer}). We note
that the entropy is oscillatory and increasing, but not monotonically. Thus, large spheres generally have greater entropies than small spheres. We also note that the oscillations are related to the presence of scattering resonances in the Mie coefficients $A_l$. 

Next, we consider a system of point-scatterers. Using (\ref{psi_s-particles}), the identity
\begin{equation}
e^{i\Bx\cdot\By}= 4\pi \sum_{l,m} i^l j_l(xy)Y_{lm}(\x)Y_{lm}^*(\y) 
\end{equation}
and the orthogonality of the spherical harmonics, we see that ${\mathcal A}^*{\mathcal A}$ can be written in the form

\begin{figure}[t]     
\centering
\includegraphics[width=3.in]{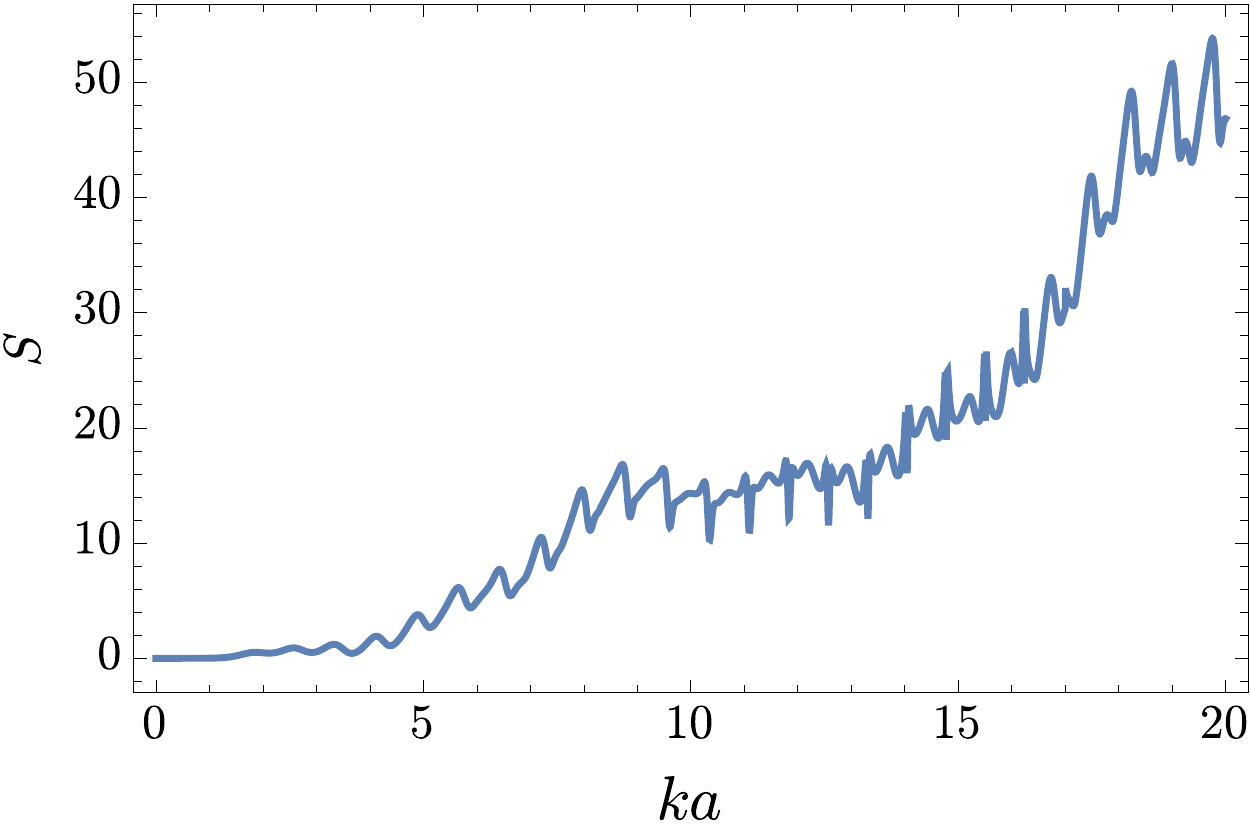} 
\caption{Entropy of entanglement of a spherical scatterer as a function of radius $a$. The index of refraction of the sphere is $n=1.5$.}
\label{fig3}
\end{figure}

\begin{equation}
\label{Psi*Psi}
\left({\mathcal A}^*{\mathcal A}\right)(\khat,\khat') = \sum_{l,m}\sum_{l', m'}A_{lm}^{l'm'}Y_{lm}(\khat)Y_{l'm'}^*(\khat') \ ,
\end{equation}
where
\begin{equation}
A_{l'm'}^{lm} = \sum_{l'',m''} C_{l''m''}^{{lm}*}C_{l'm'}^{l''m''} \ .
\end{equation}
The coefficients $C_{lm}^{l'm'}$ contain the information on the positions of the scatterers and are defined by

\begin{eqnarray}
\nonumber
C_{l'm'}^{lm} =  (4\pi)^2\sum_{\substack{{a,a'}\\{b,b'}}}i^{l+l'}
t_{ab}(k)t_{a'b'}(k'){\rm sinc}\left(|k\Br_a + k'\Br_{a'}|\right) \\
\times j_l(k r_b)j_{l'}(k'r_{b'})Y_{lm}^*(\r_a)Y_{l'm'}(\r_{a'}) \ .  
\quad\quad 
\end{eqnarray}
To construct the singular value decomposition of ${\mathcal A}$,
we expand the singular functions $u_n$ (which satisfy~(\ref{def_un})) 
into spherical harmonics of the form
\begin{equation}
u_n(\khat) = \sum_{l,m} u_{lm}^{(n)}Y_{lm}(\khat) \ ,
\end{equation}
where the coefficients $u_{lm}^{(n)}$ are to be determined.
Making use of (\ref{Psi*Psi}) and the orthogonality of the spherical harmonics,
we find that the  $u_{lm}^{(n)}$ can be obtained from the solution to the
eigenproblem
\begin{equation}
\sum_{l',m'} A_{lm}^{l'm'} u_{l'm'}^{(n)}  = \sigma_n^2 u_{lm}^{(n)} \ .
\end{equation}
Once the above eigenproblem has been solved, the entropy is computed
from (\ref{entropy}). In Fig.~4 we plot the entropy as a function of the distance
between a pair of point scatterers. We see that the entropy decreases as the separation between the scatterers increases. In the limit where the scatterers are far apart (noninteracting), the entropy vanishes, consistent with our results for the case of a single point scatterer.

\begin{figure}[b]     
\centering
\includegraphics[width=3.in]{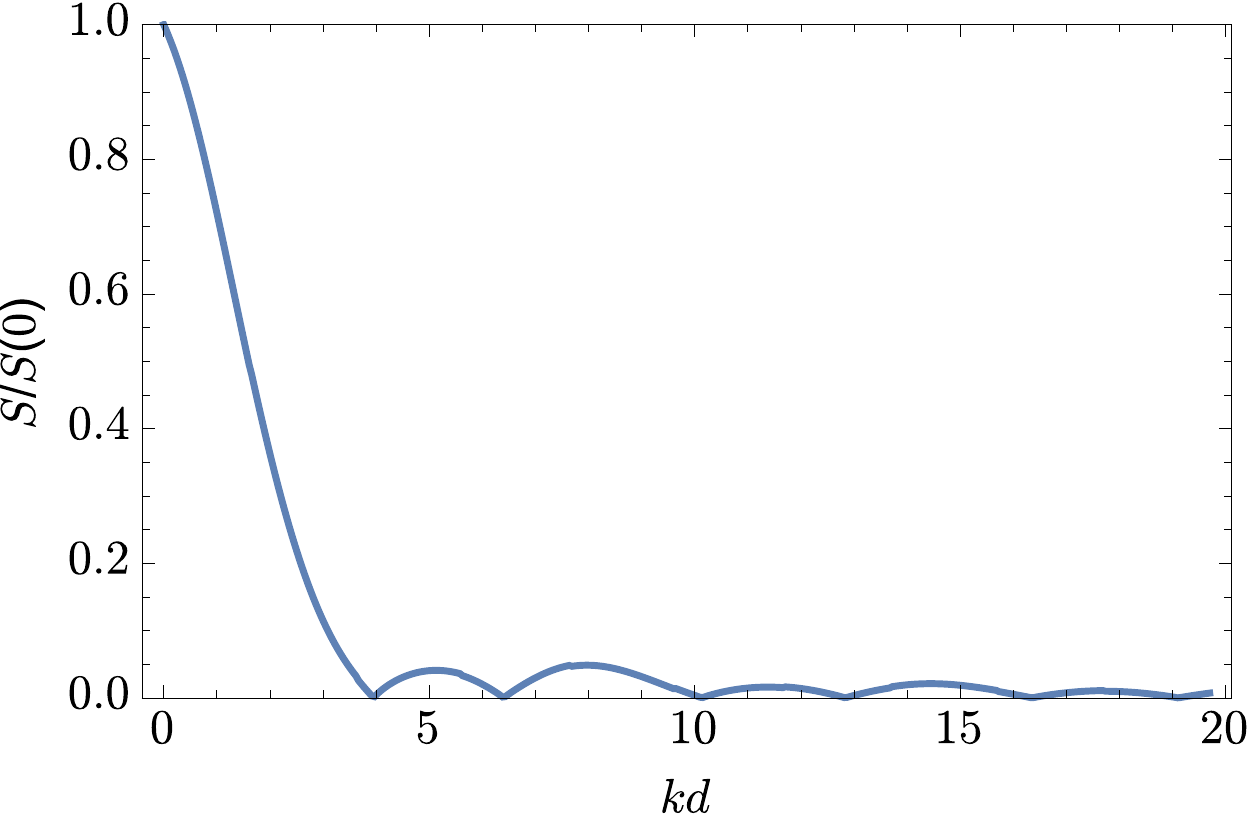} 
\caption{The entropy of entanglement of a pair of scatterers separated by a distance $d$. The radii of the scatterers is $ka=0.2$ and their index of refraction is $n=1.5$.}
\label{fig4}
\end{figure}

We close with a few remarks. (i) There is a well-known duality between partially coherent and partially entangled light~\cite{Saleh_2000}. We note that our results are analogous to the change in coherence that can occur with scattering~\cite{Mandel_Wolf}. (ii) It has been reported that entanglement can be induced by multiple scattering in random media~\cite{Ott_2010}. The  opposite conclusion was argued in \cite{Markel_2014}. Since we consider deterministic systems, the results of this Letter are potentially significant because they remove from consideration the role of randomness in modifying the entanglement of the incident field.
(iii) Although in our model the electromagnetic field is quantized, the interaction of the field with the scattering medium is treated classically. It would be of interest to extend our results to the case in which the medium consists of a collection of two-level atoms. In this manner, it should be possible to understand the transfer of entanglement from the field to the medium~\cite{Berman_2007}. 

In conclusion, we have studied the scattering of entangled two-photon states from nonabsorbing material media. In the setting of  simple model systems, we have calculated the entropy of entanglement and have characterized the influence of the entanglement of the incident field on the entanglement of the scattered field. In future work, we plan to investigate whether the observed resonances in the entropy are statistically stable in random media. 

The authors are grateful to Paul Berman for valuable discussions. This work was supported in part by the NSF Center for Photonic and Multiscale Nanomaterials under the grant DMR--1120923.

\newpage

\begin{center}
\bf Supplementary Information
\end{center}

Here we collect some basic results about point scatterers. First, consider a small spherical scatterer of radius $a$, where $ka \ll 1$. The $T$-matrix is then given by
\begin{equation}
\label{point_scatterer_again}
\bra{\Bk}T\ket{\Bk'} = t(k)e^{i(\Bk-\Bk')\cdot\Br_0}  \ ,
\end{equation}
where $t(k)=\alpha k^2$ and $\Br_0$ is the position of the scatterer. The renormalized polarizability $\alpha$ is defined by 
\begin{equation}
\alpha = \frac{\alpha_0}{1-3\alpha_0k^2/(2a) - i\alpha_0 k^3}  \ ,
\end{equation}
where the polarizability $\alpha_0=a^3(n^2-1)/3$, with $n$ the index of refraction.
Note that the above formula includes radiative corrections to the Lorentz-Lorenz form of the polarizability. Next, we consider a collection of identical point scatterers. The susceptibility is of the form $\eta(\Br)=\eta_0\sum_a \Delta(\Br-\Br_a)$, where $\eta_0=(n^2-1)/4\pi$, $\{\Br_a\}$ are the positions of the scatterers and 
\begin{equation}
\Delta(\Br) = \begin{cases}
& 1 \ , \quad   |\Br|\le a \ , \\
& 0 \ , \quad |\Br|>0 \ .
\end{cases}
\end{equation}
The $T$-matrix is given by
\begin{equation}
\label{T-multiple_again}
\bra{\Bk}T\ket{\Bk'} = \sum_{a,b}t_{ab}(k)e^{i(\Bk\cdot\Br_a-\Bk'\cdot\Br_b)} \ ,
\end{equation}
where
\begin{equation}
t_{ab} = \alpha_0 k^2M_{ab}^{-1} \ .
\end{equation}
Here
\begin{eqnarray}
M_{ab} = \delta_{ab} - \alpha_0 k^2G_{ab}  \ ,
\end{eqnarray}
where
\begin{eqnarray}
G_{ab} = \begin{cases}
& G(\Br_a,\Br_b) \ , \quad   a\ne b \ , \\
&\dfrac{3}{2a} + ik \ , \quad a=b \ .
\end{cases}
\end{eqnarray}

\end{document}